\documentclass[sigconf]{acmart}

\usepackage{booktabs} 

\usepackage{url}
\usepackage{multirow}
\usepackage{amsthm}
\usepackage{amssymb}
\usepackage{amsfonts}
\usepackage{color}
\usepackage{MnSymbol}
\usepackage{makecell}
\usepackage{arydshln}
\usepackage{graphicx}
\usepackage{microtype}
\usepackage[shortlabels]{enumitem}

\usepackage{caption} 
\setcopyright{rightsretained}

\copyrightyear{2018} 
\acmYear{2018} 
\setcopyright{acmcopyright}
\acmConference[ICMI '18]{2018 International Conference on Multimodal Interaction}{October 16--20, 2018}{Boulder, CO, USA}
\acmBooktitle{2018 International Conference on Multimodal Interaction (ICMI '18), October 16--20, 2018, Boulder, CO, USA}
\acmPrice{15.00}
\acmDOI{10.1145/3242969.3243019}
\acmISBN{978-1-4503-5692-3/18/10}

\newcommand{\ourl}{Multimodal Local-Global Ranking Fusion}
\newcommand{\ours}{MLRF}

\begin{document}
\title{Multimodal Local-Global Ranking Fusion \\ for Emotion Recognition}

\author{Paul Pu Liang}
\affiliation{
  \institution{Carnegie Mellon University}
  \city{Pittsburgh} 
  \state{PA}
  \postcode{15213}
  \country{USA}
}
\email{pliang@cs.cmu.edu}

\author{Amir Zadeh}
\affiliation{
  \institution{Carnegie Mellon University}
  \city{Pittsburgh} 
  \state{PA}
  \postcode{15213}
  \country{USA}
}
\email{abagherz@cs.cmu.edu}

\author{Louis-Philippe Morency}
\affiliation{
  \institution{Carnegie Mellon University}
  \city{Pittsburgh} 
  \state{PA}
  \postcode{15213}
  \country{USA}
}
\email{morency@cs.cmu.edu}

\title[{Multimodal Local-Global Ranking Fusion for Emotion Recognition}]{Multimodal Local-Global Ranking Fusion \\ for Emotion Recognition}

\begin{abstract}
Emotion recognition is a core research area at the intersection of artificial intelligence and human communication analysis. It is a significant technical challenge since humans display their emotions through complex idiosyncratic combinations of the language, visual and acoustic modalities. In contrast to traditional multimodal fusion techniques, we approach emotion recognition from both direct person-independent and relative person-dependent perspectives. The direct person-independent perspective follows the conventional emotion recognition approach which directly infers absolute emotion labels from observed multimodal features. The relative person-dependent perspective approaches emotion recognition in a relative manner by comparing partial video segments to determine if there was an increase or decrease in emotional intensity. Our proposed model integrates these direct and relative prediction perspectives by dividing the emotion recognition task into three easier subtasks. The first subtask involves a multimodal local ranking of relative emotion intensities between two short segments of a video. The second subtask uses local rankings to infer global relative emotion ranks with a Bayesian ranking algorithm. The third subtask incorporates both direct predictions from observed multimodal behaviors and relative emotion ranks from local-global rankings for final emotion prediction. Our approach displays excellent performance on an audio-visual emotion recognition benchmark and improves over other algorithms for multimodal fusion.
\end{abstract}

\ccsdesc[500]{Computing methodologies~Artificial Intelligence}
\ccsdesc[500]{Computing methodologies~Machine Learning}

\keywords{Multimodal Fusion; Neural Networks, Emotion Recognition}

\maketitle

\section{Introduction}
Emotion recognition is a core research area at the intersection of artificial intelligence and human communication analysis. It has immense applications towards robotics \citep{8039024,s131115549}, dialog systems \citep{Pittermann2010,Pittermann:2009:HEH:1822595}, intelligent tutoring systems \citep{Vail:2016:GDF:2930238.2930257,10.2307/jeductechsoci.18.4.435,PETROVICA2017437}, and healthcare diagnosis \citep{5373931}. Emotion recognition is multimodal in nature as humans utilize multiple communicative modalities in a structured fashion to convey emotions \citep{Busso2008IEMOCAP:Interactiveemotionaldyadic}. Two of these important modalities are acoustic and visual. In the acoustic modality, humans use prosody and various vocal expressions. In the visual modality, humans utilize facial expressions, hand gestures, and body language. Each modality is crucial when analyzing human emotions, making emotion recognition a challenging domain of artificial intelligence.

\begin{figure}[t!]
\centering{
\includegraphics[width=\linewidth]{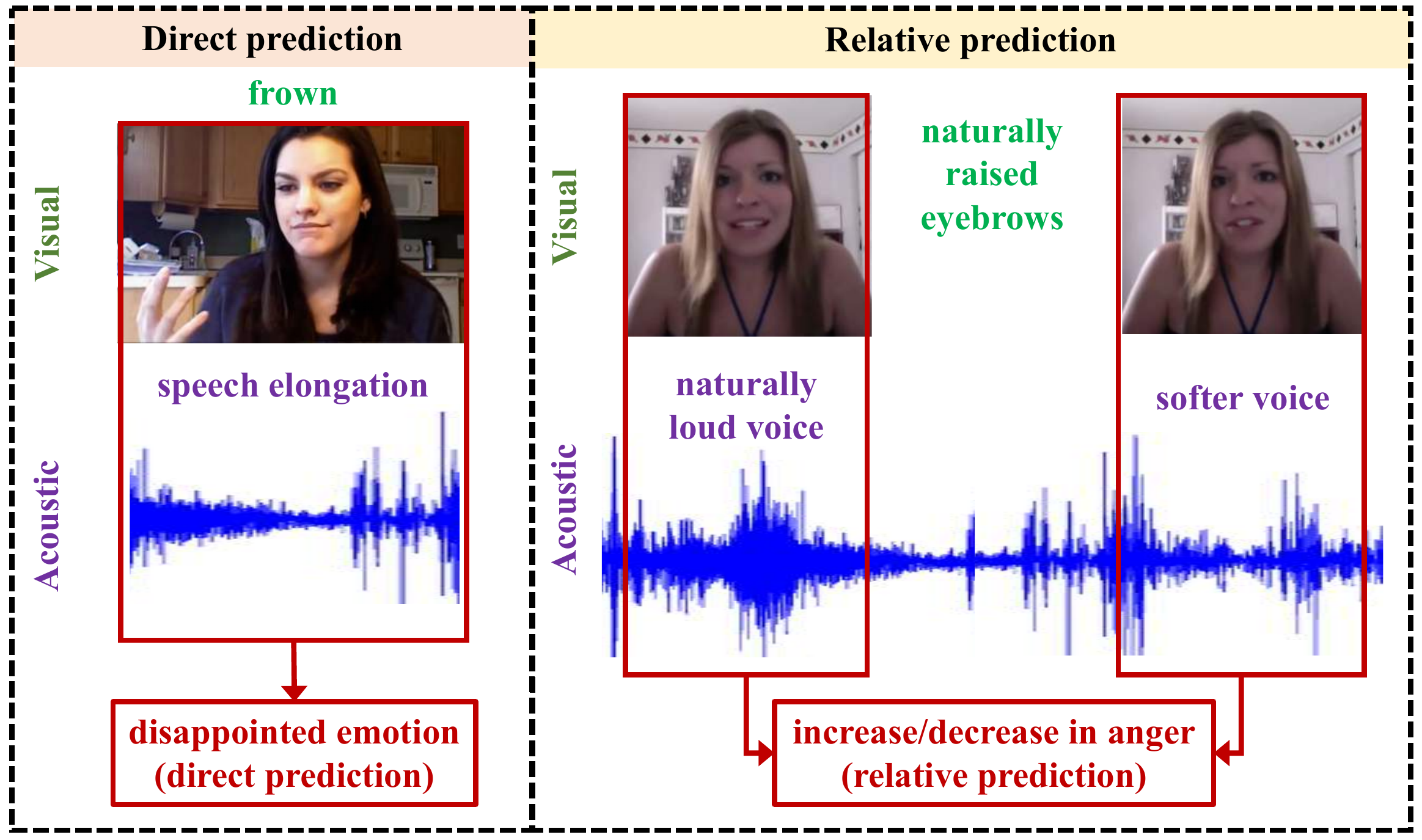}}
\caption{Our proposed Multimodal Local-Global Ranking Fusion (MLRF) model integrates both direct and relative prediction perspectives. The direct prediction perspective follows the conventional recognition approach which directly infers absolute emotion labels from observed multimodal features. The relative prediction perspective approaches emotion recognition in a relative manner by comparing partial video segments to determine changes in emotional intensity.}
\label{fig:overview}
\vspace{-4mm}
\end{figure}

Some emotional expressions are almost universal and can be recognized directly from a video segment. For example, an open mouth with raised eyebrows and a loud voice is likely to be associated with surprise. These can be seen as \textit{person-independent} behaviors and absolute emotions can be directly inferred from these behaviors (left panel of Figure \ref{fig:overview}). However, emotions are also expressed in a \textit{person-dependent} fashion with idiosyncratic behaviors. In these cases, it may not be possible to directly estimate absolute emotion intensities. Instead, it would be easier to compare two video segments of the same person and judge whether there was a relative change in emotion intensities (right panel of Figure \ref{fig:overview}). For example, a person could have naturally furrowed eyebrows and we should not always interpret this as a display of anger, but rather compare all nonverbal behaviors between two video segments to determine the relative changes in his displayed anger. From a psychological approach, research has also highlighted the advantages of using ordinal (relative) representations of human signals~\citep{cogprints730,Stewart2005AbsoluteIB,doi:10.1111/j.1460-9568.2008.06202.x}.


In this paper, we introduce the \ourl \ (\ours) model which performs emotion recognition by integrating both \textit{direct} prediction and \textit{relative} prediction approaches. This is performed by dividing the emotion recognition task into three easier multimodal subtasks (Figure~\ref{fig:model}). The first subtask is the \textit{multimodal local ranking} task. Given two short segments randomly selected from an entire video, the model is tasked with determining if there was an increase or decrease in the displayed emotion intensity. This task is often simpler than the direct emotion recognition problem since the model only needs to compare relative emotion ranks rather than compute the absolute intensities. The second subtask is the \textit{global ranking} task, which uses the previous results of local rankings to infer relative global emotion ranks using a Bayesian skill rating algorithm \citep{8273587,NIPS2006_3079}. The third subtask involves \textit{direct-relative fusion} of direct emotion predictions from observed multimodal behaviors with relative emotion ranks estimated from local-global ranking. This integration of direct and relative emotion predictions allows \ours \ to model both person-independent and person-dependent behaviors for complete emotion recognition. We show that \ours \ is suitable for multimodal tasks by performing experiments on an audio-visual emotion recognition benchmark. The proposed \ours \ approach displays excellent performance over the baselines, improving over other algorithms for multimodal fusion.


\section{Related Work}
\label{Related Work}

Previous approaches in multimodal emotion recognition can be categorized as follows:

\noindent \textbf{Non-temporal Models:} These approaches simplify the temporal aspect by averaging information through time~\cite{abburi2016multimodal}. Fusion is performed by concatenating the multimodal inputs. However, these methods tend to overfit without discovering generalizable speaker-independent and speaker-dependent features~\cite{xu2013survey}. More complex fusion methods learn separate models for each modality and combine the outputs~\cite{Snoek:2005:EVL:1101149.1101236}. However, simple decision voting is unable to discover the complex multimodal combinations involved in speaker-dependent features~\citep{multistage}.



\noindent \textbf{Temporal Models:} Long Short-term Memory Networks (LSTMs) \cite{hochreiter1997long,6638947,Schuster:1997:BRN:2198065.2205129} have been extended for multimodal settings~\cite{rajagopalan2016extending} and with binary gating mechanisms to remove noisy modalities~\cite{Chen:2017:MSA:3136755.3136801}. More advanced models use memory mechanisms~\cite{zadeh2018memory}, low-rank approximations to tensor products~\cite{lowrank}, multiple attention stages~\cite{multistage} or assignments~\cite{zadeh2018multi} or generative-discriminative objectives to learn factorized~\cite{factorized} or joint multimodal representations~\citep{seq2seq}. To our knowledge, our approach is the first to approach multimodal fusion with a neural local-global ranking fusion approach. The strength of our approach lies in approaching both speaker-independent and speaker-dependent features via direct and relative emotion predictions respectively. Algorithmically, our divide-and-conquer insight simplifies the emotion recognition task into three easier multimodal subtasks and allows us to incorporate probabilistic structure as compared to entirely neural approaches. 

Our work is also related to ranking algorithms. Bayesian ranking algorithms have been used in ranking the skills of players in Chess~\cite{elo1978rating,6933058} and online games~\cite{NIPS2006_3079}. Recently, ranking methods have been applied for facial expression intensity estimation~\cite{8273587}. To the best of our knowledge, we are the first to integrate relative measures from local-global ranking with direct predictions for emotion recognition. We also apply our approach to a multimodal setting where temporal information is primordial and there exist complex interactions between the acoustic and visual modalities.

\section{\ourl}

\begin{figure}[t!]
\centering{
\includegraphics[width=\linewidth]{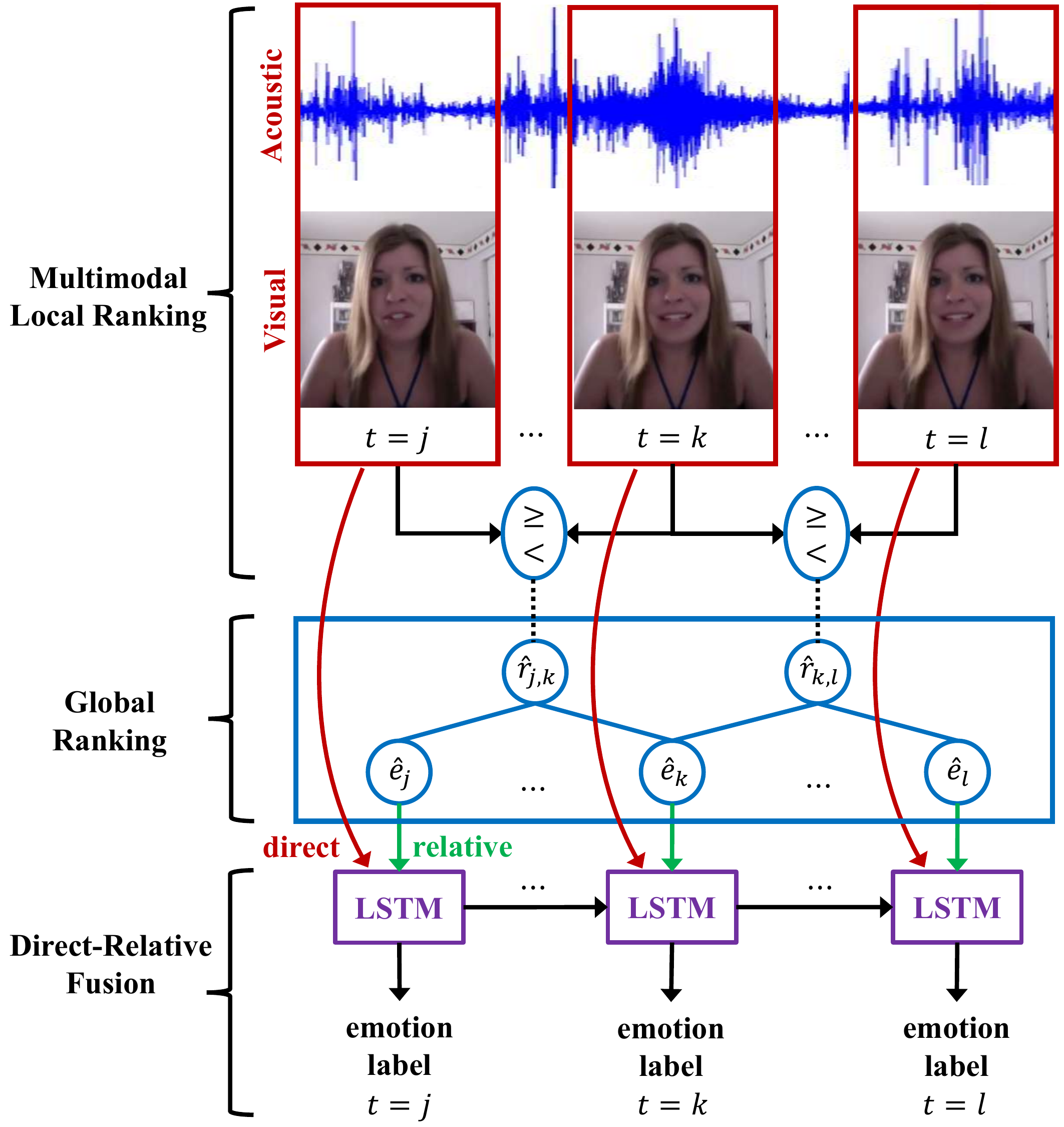}}
\caption{In \ours, the emotion recognition task is divided into three easier subtasks: (1) \textit{multimodal local ranking} involves a local ranking of emotion intensities between two short segments of a video, (2) \textit{global ranking} uses the results of local rankings to infer global relative emotion ranks using a Bayesian ranking algorithm, and (3) \textit{direct-relative fusion} integrates direct emotion predictions estimated from observed multimodal behaviors with relative emotion ranks from local-global rankings for final emotion recognition.}
\label{fig:model}
\vspace{-4mm}
\end{figure}

The \ourl \ (\ours) model (Figure~\ref{fig:model}) aims to integrate both direct and relative emotion prediction approaches to model person-independent and person-dependent behaviors. This is achieved by subdividing the emotion recognition task into three easier subtasks: (1) \textit{multimodal local ranking}, (2) \textit{global ranking}, and (3) \textit{direct-relative fusion}. Relative emotion rankings are inferred from the first two sub-tasks. The third subtask performs direct emotion predictions while at the same time integrating relative emotion rankings for final emotion recognition.

\subsection{Problem Statement}

Given a set of modalities $\mathcal{M}$ each in the form of a temporal sequence with $T$ time steps, we denote the data from modality $m \in \mathcal{M}$ as $\mathbf{x}^m_{} = \langle {\mathbf{x}_{1}^m}_{}, {\mathbf{x}_{2}^m}_{}, \cdots, {\mathbf{x}_{T}^m}_{} \rangle$, where ${\mathbf{x}^m_{t}}_{} \in \mathbb{R}^{d_m}$ denotes the input of modality $m$ at time $t$ with dimensionality $d_m$. For a window size $w$, define $\mathbf{x}^m_{t_w} = \langle \mathbf{x}_{t-w}^m, \cdots \mathbf{x}_{t}^m, \cdots \mathbf{x}_{t+w}^m \rangle$ as the short video segment centered around time $t$ with a time window $w$. The goal is to estimate the sequence of emotion labels $\mathbf{y}_{} = \langle {y_1}_{}, {y_2}_{}, \cdots, {y_T}_{} \rangle$, where ${y_t}_{} \in \mathbb{R}$ is the emotion label at time $t$. In our experiments, the training dataset $\mathcal{D}$ consists of $n$ input-label pairs $\mathcal{D} = \{ \mathbf{x}^m_{(i)}, \mathbf{y}_{(i)} : m \in \mathcal{M} \}_{i=1}^n$. The test dataset has a similar structure but with no overlapping speakers. The subscript ${(i)}$ indicates the variable associated with the $i$-th video. We drop index ${(i)}$ when it is clear from the context.

\subsection{Multimodal Local Ranking}

In the \textit{multimodal local ranking task}, the model is presented with two short segments randomly selected within a video and is tasked with determining whether there was an increase or decrease in emotion intensity. This local ranking process is repeated multiple times for different segment pairs. The number of pairs is a hyper-parameter and $(J,K)$ denotes the set of all pairs. Given a pair $(j,k) \in (J,K)$, $ \mathbf{x}_{j_w} = \{ \mathbf{x}^m_{j_w} : m \in \mathcal{M} \}$ denotes a short segment integrating all multimodal features and likewise for $ \mathbf{x}_{k_w}$. $w$ is a hyper-parameter which defines the context over which local ranking is performed. The local rank $r_{j,k} = \mathbb{I}[y_j > y_k]$ indicates whether there was an increase or decrease of emotion intensity between segments $j$ and $k$. The multimodal local ranking dataset $\mathcal{D}_{local}$ is therefore:
\begin{equation}
\mathcal{D}_{local} = \{ \{ {\mathbf{x}_{j_w}}_{(i)}, {\mathbf{x}_{k_w}}_{(i)}, {r_{j,k}}_{(i)} \}_{j \in J_{(i)}, k \in K_{(i)}} \}_{i=1}^n
\end{equation}
This defines a binary classification problem over two short segments. This task is simpler than the direct emotion recognition problem since the local ranking model only needs to compare the differences in emotion intensities rather than compute the exact emotion intensities themselves.

To solve the multimodal local ranking problem, we define an estimator $f_{local}$ with parameters $\Theta_{local}$:
\begin{equation}
\hat{r}_{j,k} = f_{local}(\mathbf{x}_{j_w},\mathbf{x}_{k_w}\ ; \ \Theta_{local})
\end{equation}
Solving this problem involves multimodal fusion since $\mathbf{x}_{j_w}$ and $\mathbf{x}_{k_w}$ represent data from the input modalities $\mathcal{M}$. In order to solve for parameters $\Theta_{local}$, we minimize the empirical measure of the categorical cross-entropy between the target local ranks ${r}_{j,k}$ and our estimated local ranks $\hat{r}_{j,k}$:
\begin{equation}
\hat{\mathcal{L}}_{local} = \frac{1}{|\mathcal{D}_{local}|} \sum_{ (\mathbf{x}_{j_w},\mathbf{x}_{k_w}, r_{j,k})\in \mathcal{D}_{local}} {r}_{j,k} \log \hat{r}_{j,k}
\end{equation}

In practice, we parametrize $f_{local}$ using an LSTM \citep{hochreiter1997long} which takes as input the differences of multimodal tensors:
\begin{equation}
\mathbf{x}_{local} = \bigoplus_{m \in \mathcal{M}} \mathbf{x}^m_{k_w} - \bigoplus_{m \in \mathcal{M}} \mathbf{x}^m_{j_w}
\end{equation}
where $\bigoplus$ denotes tensor concatenation of all unimodal feature vectors $\mathbf{x}^m$. $\mathbf{x}_{local}$ is the input sequence to the LSTM that performs multimodal local ranking. A neural network classification layer on the final LSTM output $\mathbf{h}_T$ is used to estimate local ranks $\hat{r}_{j,k}$.

In contrast to \citep{8273587} who developed a model for images only, our problem involves multimodal video segments. In our models, local comparisons of emotion intensities are performed over a time window $w$. Emotion intensities often require more than still frames, especially when including the acoustic modality. Furthermore, human communicative behaviors can be asynchronous and a longer time window is required to track changes in both visual and acoustic behaviors displayed by the person.

\subsection{Global Ranking}

The second task is the \textit{global ranking} task, which uses the previous results of local rankings to infer global emotion ranks using a Bayesian skill rating algorithm \citep{NIPS2006_3079}. The algorithm will infer global emotion ranks $e_t$ at each time step $t$ of the multimodal video. These ranks are initially sampled from a prior distribution $\mathcal{N}(\mu_t,\sigma_t^2)$. The algorithm models global emotion ranks $e_t$ as hidden variables that are not directly observed from the data. What we do observe are local rankings ${r}_{j,k}$ between two time segments $j$ and $k$. By our definition of local ranks, the conditional probabilities of local ranks $p({r}_{j,k}=1|e_j,e_k)$ will be equal to $p(e_j > e_k)$, the probability of rank $e_j$ being larger than $e_k$. Given estimated ranks $\hat{r}_{j,k}$, we estimate global relative emotion ranks $\hat{e}_t$ using a ranking algorithm which involves message passing over factor graph models. A detailed treatment is presented in \citep{NIPS2006_3079,8273587}. After all iterations of global ranking, we obtain estimated global emotion ranks $\hat{e}_t$ for all video segments.

\subsection{Direct-Relative Fusion}

The third task involves \textit{direct-relative fusion}. The global emotion ranks $\hat{\mathbf{e}} = \langle \hat{e}_1, \cdots , \hat{e}_T \rangle $ are incorporated with the raw multimodal inputs ${\mathbf{x}^m}$ to estimate final emotion intensities. This allows us to perform \textit{direct} estimation of the absolute emotion intensities from multimodal data while at the same time integrating \textit{relative} emotion ranks from local-global rankings.


The integration of direct and relative predictions is performed by learning a function $f_{fusion}$ with parameters $\Theta_{fusion}$:
\begin{equation}
\hat{\mathbf{y}} = f_{fusion} (\mathbf{x}, \hat{\mathbf{e}} \ ; \ \Theta_{fusion})
\end{equation}
where $\hat{\mathbf{y}}$ are the predicted emotion labels. We solve for $\Theta_{fusion}$ by minimizing an empirical measure of the loss between $\hat{\mathbf{y}}$ and $\mathbf{y}$:
\begin{equation}
\hat{\mathcal{L}}_{fusion} = \frac{1}{n} \sum_{i=1}^n \ell (\hat{\mathbf{y}}_{(i)}, {\mathbf{y}}_{(i)} ) 
\end{equation}
where $\ell (\hat{\mathbf{y}}, {\mathbf{y}})$ is a loss function over two segments of emotion intensities. In practice, we parametrize $f_{fusion}$ with an LSTM on the concatenated multimodal data and global emotion ranks:
\begin{equation}
\mathbf{x}_{fusion} = \bigg( \bigoplus_{m \in \mathcal{M}} \mathbf{x}^m \bigg) \bigoplus \hat{\mathbf{e}}
\end{equation}
where $\bigoplus$ denotes tensor concatenation. Using an LSTM on $\mathbf{x}_{fusion}$ allows us to capture the temporal dependencies across multimodal time series data and label emotion intensities. A time-distributed neural network regression layer is used on the LSTM outputs $\mathbf{h}_t$ to estimate final emotion intensities $\hat{y}_t$.
The global emotion ranks $\hat{\mathbf{e}}$ can be incorporated with multimodal data $\mathbf{x}^m$ using any multimodal fusion method. As a result, our approach represents a generalizable framework to integrate relative emotion intensities into a variety of multimodal fusion models.


\section{Experiments}

\subsection{Dataset}

We use the AVEC16 dataset (RECOLA) \citep{conf/fgr/RingevalSSL13} for audio-visual emotion recognition. AVEC16 consists of 9 training videos, 8 are used to optimize parameters (our training set) and 1 is held out to tune hyperparameters (our validation set). The 9 validation videos are used as our test set to compare each method (test videos are not publicly released). We use the provided appearance and geometric visual features and acoustic features. Each video has 7501 time steps after alignment between the modalities and is labeled continuously for arousal and valence at every time step. The metric used is the concordance correlation coefficient (CCC) \citep{article}. 


\subsection{Baseline Models}

We compare to the following: \textbf{EF-LSTM} (Early Fusion LSTM) uses a single LSTM \cite{hochreiter1997long} on concatenated multimodal inputs. We also implement the \textbf{EF-SLSTM} (stacked) \citep{6638947}, \textbf{EF-BLSTM} (bidirectional) \citep{Schuster:1997:BRN:2198065.2205129} and \textbf{EF-SBLSTM} (stacked bidirectional) versions and report the best result. \textbf{GF-LSTM} (Gated Fusion LSTM) \citep{Yu:2017:TSA:3123266.3123413} extends the EF-LSTM by assigning an LSTM to each modality and combining the final LSTM outputs with a gated attention fusion layer for final prediction. \textbf{MV-LSTM} (Multi-View LSTM) \cite{rajagopalan2016extending} allocates regions inside a LSTM to different modalities using parameters $\alpha$ and $\beta$. We experiment with the view-specific ($0 < \alpha \le 1, \beta=0$), coupled ($\alpha=0, 0 < \beta \le 1$), hybrid ($ 0 < \alpha < 1, 0 < \beta < 1$) and fully connected ($\alpha=1, \beta=1$) topologies as well. Our model is indicated by \textbf{\ours-$\mathbf{k}$} where $k$ is the number of local comparison pairs. We set the default window size for local ranking as $w=200$.

\subsection{Results on Emotion Recognition}

\newcolumntype{K}[1]{>{\centering\arraybackslash}p{#1}}
\definecolor{gg}{RGB}{45,190,45}

\begin{table}[t!]
\fontsize{7.5}{10}\selectfont
\setlength\tabcolsep{1.5pt}
\begin{tabular}{l : *{2}{K{1.3cm}}}
\Xhline{3\arrayrulewidth}
Dataset & \multicolumn{2}{c}{\textbf{AVEC16}} \\
Task & \multicolumn{1}{c}{Arousal} & \multicolumn{1}{c}{Valence} \\
Metric       & CCC & CCC \\ 
\Xhline{0.5\arrayrulewidth}
EF-(-/S/B/SB)LSTM	\citep{hochreiter1997long,6638947,Schuster:1997:BRN:2198065.2205129}	& 0.4327 & 0.4667 \\
Gated-LSTM \citep{Yu:2017:TSA:3123266.3123413}	& 0.3210 & 0.4667 \\
MV-LSTM, view-specific \cite{rajagopalan2016extending} & 0.4530 & 0.4431 \\
MV-LSTM, coupled \cite{rajagopalan2016extending} & 0.4300 & 0.4477 \\ 
MV-LSTM, hybrid \cite{rajagopalan2016extending} & 0.4729 & 0.4924 \\
MV-LSTM, fully connected \cite{rajagopalan2016extending} & 0.4293 & 0.4896 \\
\Xhline{0.5\arrayrulewidth}
{\ours-500}		& 0.4732 & 0.5063 \\
{\ours-1000}	& \textbf{0.5049} & \textbf{0.5432} \\
\Xhline{0.5\arrayrulewidth}
Improvement over baselines	& \textcolor{gg}{$\uparrow $ \textbf{0.032}} & \textcolor{gg}{$\uparrow $ \textbf{0.0508}} \\ 
\Xhline{3\arrayrulewidth}
\end{tabular}
\caption{Multimodal emotion recognition results on AVEC16 dataset. The best results are highlighted in bold and green indicates the improvement over baselines. The \ours \ outperforms the baselines across all evaluation metrics.}
\label{avec16}
\vspace{-4mm}
\end{table}

\ours \ achieves better results on arousal and valence regression as compared to the baselines (Table~\ref{avec16}). Our results show that incorporating local-global ranking estimates into simple models (EF-LSTM in our experiments) prove more effective than engineering complex neural architectures such as the Gated-LSTM~\citep{Yu:2017:TSA:3123266.3123413} and the MV-LSTM~\citep{rajagopalan2016extending}. Furthermore, although the videos contain more than 7500 time steps, simply sampling 500-1000 local ranking pairs per video significantly improved final performance. As a result, incorporating relative emotion intensities via \ours \ is an effective method without adding significant computational complexity.

\subsection{Discussion}

\newcolumntype{K}[1]{>{\centering\arraybackslash}p{#1}}
\definecolor{gg}{RGB}{45,190,45}

\begin{table}[t!]
\fontsize{7.5}{10}\selectfont
\setlength\tabcolsep{1.5pt}
\begin{tabular}{l : *{2}{K{1.3cm}}}
\Xhline{3\arrayrulewidth}
Dataset & \multicolumn{2}{c}{\textbf{AVEC16}} \\
Task & \multicolumn{1}{c}{Arousal} & \multicolumn{1}{c}{Valence} \\
Metric       & CCC & CCC \\ 
\Xhline{0.5\arrayrulewidth}
{\ours-500 $w=10$} &  0.4165 & 0.2377 \\
{\ours-500 $w=50$} & 0.4168 & 0.4175 \\
{\ours-500 $w=100$} &  0.4196 & 0.4340 \\
{\ours-500 $w=200$}	&  \textbf{0.4732} & \textbf{0.5063} \\
\Xhline{3\arrayrulewidth}
\end{tabular}
\caption{Increasing the window size $w$ improves performance. We observed a similar trend for \ours-1000.}
\label{window}
\vspace{-4mm}
\end{table}

\newcolumntype{K}[1]{>{\centering\arraybackslash}p{#1}}
\definecolor{gg}{RGB}{45,190,45}

\begin{table}[t!]
\fontsize{7.5}{10}\selectfont
\setlength\tabcolsep{1.5pt}
\begin{tabular}{l : *{2}{K{1.3cm}}}
\Xhline{3\arrayrulewidth}
Dataset & \multicolumn{2}{c}{\textbf{AVEC16}} \\
Task & \multicolumn{1}{c}{Arousal} & \multicolumn{1}{c}{Valence} \\
Metric       & CCC & CCC \\ 
\Xhline{0.5\arrayrulewidth}
{\ours-500 direct predictions only} & 0.4327 & 0.4667 \\
{\ours-500 relative predictions only} & 0.3646 & 0.0402 \\
{\ours-500}	& \textbf{0.4732} & \textbf{0.5063} \\
\Xhline{0.5\arrayrulewidth}
{\ours-1000 direct predictions only} & 0.4327 & 0.4667 \\
{\ours-1000 relative predictions only}	&  0.4297 & 0.0846 \\
{\ours-1000}	& \textbf{0.5049} & \textbf{0.5432} \\
\Xhline{3\arrayrulewidth}
\end{tabular}
\caption{Ablation studies: incorporating both direct and relative emotion predictions is crucial for performance.}
\label{ablation}
\vspace{-4mm}
\end{table}

\noindent \textbf{Effect of Number of Local Comparison Pairs:} Table~\ref{avec16} shows that performance increases as the number of sampled local comparison pairs increases. More observations of local ranks $\hat{r}_{j,k}$ improves our estimates of global emotion ranks $\hat{e}_t$, which in turn provide better relative emotion intensities for emotion recognition.

\noindent \textbf{Effect of Window Size:} From Table~\ref{window}, we observe that increasing the window size $w$ for multimodal local ranking is important. This supports the fact that human communicative behaviors are asynchronous and a longer time window is required to track changes in visual and acoustic behaviors displayed by the speaker.

\noindent \textbf{Effect of Direct and Relative Approaches:} We found that fusing direct emotion predictions from observed multimodal behaviors with relative emotion predictions from local-global ranking estimates was crucial (Table~\ref{ablation}). Therefore, integrating both direct person-independent and relative person-dependent approaches is important for emotion recognition.



\section{Conclusion}
This paper approached multimodal emotion recognition from both direct person-independent and relative person-dependent perspectives. Our proposed Multimodal Local-Global Ranking Fusion (MLRF) model integrates direct and relative predictions by dividing emotion recognition into three easier subtasks: \textit{multimodal local ranking}, \textit{global ranking} and \textit{direct-relative fusion}. Our experiments showed that \ours \ displays excellent performance on multimodal tasks. Therefore, incorporating direct emotion predictions from multimodal behaviors and relative emotion ranks from local-global rankings is a promising direction for multimodal machine learning.

\section{Acknowledgements}
This material is based upon work partially supported by Samsung. Any opinions, findings, and conclusions or recommendations expressed in this material are those of the author(s) and do not necessarily reflect the views of Samsung, and no official endorsement should be inferred.

\bibliographystyle{ACM-Reference-Format}
\bibliography{sample-bibliography}

\end{document}